# Fair Isaac Technical Paper

**Subject:** Roughness Penalty for Liquid Scorecards

**From:** Bruce Hoadley (BCH – A34  X27051)

**Date:** March 7, 2003

# Abstract

A liquid scorecard has liquid characteristics, for which the characteristic score is a smooth function of the characteristic over a liquid range. The smooth function is based on B-splines – typically cubic. In contrast, the characteristic scores for traditional scorecards are step functions of the characteristics. Previously, there were two ways to control the smoothness of the liquid characteristic score: (i) coarse classing – the fewer the number of classes, the smoother the curve, (ii) the penalty parameter, which penalizes the norm of the score coefficient vector. However, in classical cubic spline fitting theory, a direct measure of curve roughness is used as a penalty term in the fitting objective function. In this paper, I work out the details of this concept for our characteristic scores, which are linear functions of a cubic B-spline basis. The roughness penalty is the integral of the second derivative squared. As you vary the characteristic smoothness parameter from zero to infinity, the characteristic score goes from being rough to being very smooth. As one moves from rough to smooth, the palatable characteristic score jumps off the page. This is illustrated by a case study. This case study also shows that smoothness parameters, which maximize validation divergence, do not always yield the most palatable model.

# Table of Contents

# 1. Introduction

Reference [1] describes the theory behind the liquid scorecard. A liquid scorecard is a Generalized Additive Model (GAM), with a score formula of the form

$$Score = \sum_{k=1}^{p} CS_k(x_k),$$

where

$$x_k = \text{Characteristic } k$$
$$CS_k(x_k) = \text{Characteristic Score } k.$$

In a liquid scorecard, some of the characteristic scores are smooth functions of the characteristic over a liquid range of the characteristic. Over it's liquid range, a characteristic score has the form

$$CS(x) = \sum_{i=1}^{q} \beta_i B_i(x), \qquad (1)$$

where the $B_i(x)$'s are from a B-spline basis. The most popular type of B-spline is the cubic B-spline.

When fitting this type of model, it is possible for the fitted cubic spline to be rougher than palatability would dictate. The classical solution to this problem is to include in the fitting objective function a term, which penalizes roughness in the fitted function. This tends to produce a smoother solution. The classical roughness penalty term is of the form

$$\int [CS''(x)]^2 dx, \qquad (2)$$

where the integral is taken over the liquid range.

The purpose of this paper is to derive a formula for this penalty term for the characteristic score based on the cubic B-splines.

## 2. Review of B-Spline Theory

### Basis function

Here I review some of the material in Reference [1].

Associated with the spline basis functions is a set of knots, $\{k(1), k(2), ..., k(m)\}$.

The cubic spline basis functions are defined iteratively. The first step in this iteration is to define a sequence of numbers, $\{t(1), t(2), ..., t(m+6)\}$, which are related to the knots as follows:

$$t(1) = t(2) = t(3) = t(4) = k(1)$$
$$t(i) = k(i-3) \quad \text{for } i = 5, ..., m+2$$
$$t(m+3) = t(m+4) = t(m+5) = t(m+6) = k(m).$$

Next I define $4 \times (m+2)$ functions of $x$

$$B(x|i, j) \quad i = 1, 2, ..., m+2 \quad j = 1, 2, 3, 4.$$

For $j = 1$

$$B(x|i,1) = 1\{t(i) \leq x < t(i+1)\} \quad i = 1, 2, ..., m+1$$
$$B(x|m+2,1) = 1\{t(m+2) \leq x \leq t(m+3)\},$$

where $1\{\text{Event}\}$ is the indicator variable of Event.

For $j = 2,3,4$

For $i = 1,2,\ldots,m+2$

$$T_1(x|i,j) = \begin{cases} \dfrac{[x-t(i)]}{[t(i+j-1)-t(i)]} B(x|i,j-1) & \text{if } t(i+j-1)-t(i) > 0 \\ 0 & \text{otherwise} \end{cases}$$

$$T_2(x|i,j) = \begin{cases} \dfrac{[t(i+j)-x]}{[t(i+j)-t(i+1)]} B(x|i+1,j-1) & \text{if } t(i+j)-t(i+1) > 0 \\ 0 & \text{otherwise} \end{cases}$$

$$B(x|i,j) = T_1(x|i,j) + T_2(x|i,j). \tag{3}$$

Some of the functions defined above are vacuous in that they are identically equal to zero. They are

$$B(x|1,1) = B(x|2,1) = B(x|3,1) \equiv 0$$
$$B(x|1,2) = B(x|2,2) \equiv 0$$
$$B(x|1,3) \equiv 0.$$

The functions $B(x|4,1), B(x|5,1), \ldots, B(x|m+2,1)$ form a first order B-spline basis with knots, $\{k(1),\ldots,k(m)\}$. These $m-1$ basis functions are just the attribute indicator variables for the attributes defined by the knots.

The functions $B(x|1,4), B(x|2,4), \ldots, B(x|m+2,4)$ form a fourth order B-spline basis with knots, $\{k(1),\ldots,k(m)\}$. This means that every cubic spline with these knots can be expressed as a linear combination of $B(x|1,4), B(x|2,4), \ldots, B(x|m+2,4)$. These cubic splines have matching first and second derivatives at the internal knots, $\{k(2),\ldots,k(m-1)\}$.

### Derivative of a B-spline basis function

According to formula (8) on page 115 of Reference [2], the derivative of a B-spline basis function is

$$B'(x|i,j) = (3)\left[ \dfrac{-B(x|i+1,j-1)}{t(i+j)-t(i+1)} + \dfrac{B(x|i,j-1)}{t(i+j-1)-t(i)} \right]. \tag{4}$$

## 3. General Formula for the Penalty Term

Plugging formula (1) into formula (2) yields

$$\int [CS''(x)]^2 dx = \int \left[ \sum_{i=1}^{m+2} \beta_i B''(x|i,4) \right]^2 dx$$

$$= \int \sum_{i=1}^{m+2} \sum_{j=1}^{m+2} \beta_i B''(x|i,4) \beta_j B''(x|j,4) dx$$

$$= \sum_{i=1}^{m+2} \sum_{j=1}^{m+2} \beta_i \beta_j \int B''(x|i,4) B''(x|j,4) dx$$

$$= \sum_{i=1}^{m+2} \sum_{j=1}^{m+2} \beta_i \beta_j R_{ij}$$

$$= \beta' R \beta,$$

where

$$\beta' = (\beta_1, \ldots, \beta_{m+2})$$
$$R = \text{Matrix with } ij \text{ element } R_{ij}$$
$$R_{ij} = \int B''(x|i,4) B''(x|j,4) dx. \quad (5)$$

**Formula for** $B''(x|i,4)$

An application of formula (4) yields

$$B'(x|i,4) = (3) \left[ \frac{-B(x|i+1,3)}{t(i+4) - t(i+1)} + \frac{B(x|i,3)}{t(i+3) - t(i)} \right].$$

Hence

$$B''(x|i,4) = (3) \left[ \frac{-B'(x|i+1,3)}{t(i+4) - t(i+1)} + \frac{B'(x|i,3)}{t(i+3) - t(i)} \right].$$

Another application of formula (4) yields

$$B''(x|i,4) = -\frac{3}{t(i+4)-t(i+1)}(2)\left[\frac{-B(x|i+2,2)}{t(i+4)-t(i+2)} + \frac{B(x|i+1,2)}{t(i+3)-t(i+1)}\right]$$

$$+ \frac{3}{t(i+3)-t(i)}(2)\left[\frac{-B(x|i+1,2)}{t(i+3)-t(i+1)} + \frac{B(x|i,2)}{t(i+2)-t(i)}\right]$$

$$= +\frac{6}{[t(i+3)-t(i)][t(i+2)-t(i)]}B(x|i,2)$$

$$- \frac{6}{[t(i+3)-t(i)][t(i+3)-t(i+1)]}B(x|i+1,2)$$

$$- \frac{6}{[t(i+4)-t(i+1)][t(i+3)-t(i+1)]}B(x|i+1,2)$$

$$+ \frac{6}{[t(i+4)-t(i+1)][t(i+4)-t(i+2)]}B(x|i+2,2).$$

We can write this as

$$B''(x|i,4) = c_i B(x|i,2)$$
$$+ (d_i + e_i) B(x|i+1,2)$$
$$+ f_i B(x|i+2,2), \qquad (6)$$

where

$$c_i = +\frac{6}{[t(i+3)-t(i)][t(i+2)-t(i)]}$$

$$d_i = -\frac{6}{[t(i+3)-t(i)][t(i+3)-t(i+1)]}$$

$$e_i = -\frac{6}{[t(i+4)-t(i+1)][t(i+3)-t(i+1)]}$$

$$f_i = +\frac{6}{[t(i+4)-t(i+1)][t(i+4)-t(i+2)]}. \qquad (7)$$

Fair, Isaac and Company, Inc.The variables $c_i, d_i, e_i, f_i$ are defined for $i = 1, \ldots, m+2$. Whenever the denominator of one of these variables is equal to 0, the variable is set equal to 0. Hence

$$c_i = 0 \text{ for } i = 1, 2$$
$$d_i = 0 \text{ for } i = 1, m+2$$
$$e_i = 0 \text{ for } i = 1, m+2$$
$$f_i = 0 \text{ for } i = m+1, m+2.$$

Application of formula (3) to formula (6) yields

$$B''(x|i,4) = c_i \left\{ \left[ \frac{x - t(i)}{t(i+1) - t(i)} \right] B(x|i,1) + \left[ \frac{t(i+2) - x}{t(i+2) - t(i+1)} \right] B(x|i+1,1) \right\}$$
$$+ (d_i + e_i) \left\{ \left[ \frac{x - t(i+1)}{t(i+2) - t(i+1)} \right] B(x|i+1,1) + \left[ \frac{t(i+3) - x}{t(i+3) - t(i+2)} \right] B(x|i+2,1) \right\}$$
$$+ f_i \left\{ \left[ \frac{x - t(i+2)}{t(i+3) - t(i+2)} \right] B(x|i+2,1) + \left[ \frac{t(i+4) - x}{t(i+4) - t(i+3)} \right] B(x|i+3,1) \right\} \quad (8)$$

Next define the linear functions

$$P(x|i) = \begin{cases} \dfrac{x - t(i)}{t(i+1) - t(i)} & \text{if } t(i+1) - t(i) > 0 \\ 0 & \text{otherwise} \end{cases}$$

$$N(x|i) = \begin{cases} \dfrac{t(i+1) - x}{t(i+1) - t(i)} & \text{if } t(i+1) - t(i) > 0 \\ 0 & \text{otherwise}. \end{cases} \quad (9)$$

Plugging these definitions into formula (8) yields

$$B''(x|i,4) = c_i \{P(x|i)B(x|i,1) + N(x|i+1)B(x|i+1,1)\}$$
$$+ (d_i + e_i)\{P(x|i+1)B(x|i+1,1) + N(x|i+2)B(x|i+2,1)\}$$
$$+ f_i\{P(x|i+2)B(x|i+2,1) + N(x|i+3)B(x|i+3,1)\}$$
$$= [(c_i)P(x|i)]B(x|i,1)$$
$$+ [(d_i + e_i)P(x|i+1) + (c_i)N(x|i+1)]B(x|i+1,1)$$
$$+ [(f_i)P(x|i+2) + (d_i + e_i)N(x|i+2)]B(x|i+2,1)$$
$$+ [(f)_i N(x|i+3)]B(x|i+3,1). \quad (10)$$

This can be expressed in a form more suitable for computation as follows:

$$B''(x|i,4) = \sum_{k=1}^{m+5} L(x|i,k) B(x|k,1), \tag{11}$$

where

$$L(x|i,k) = a_{ik} P(x|k) + b_{ik} N(x|k)$$

$$a_{ik} = \begin{cases} 0 & k = 1,\ldots,i-1 \\ c_i & k = i \\ d_i + e_i & k = i+1 \\ f_i & k = i+2 \\ 0 & k = i+3 \\ 0 & k = i+4,\ldots,m+5 \end{cases}$$

$$b_{ik} = \begin{cases} 0 & k = 1,\ldots,i-1 \\ 0 & k = i \\ c_i & k = i+1 \\ d_i + e_i & k = i+2 \\ f_i & k = i+3 \\ 0 & k = i+4,\ldots,m+5. \end{cases} \tag{12}$$

**Formula for $R_{ij}$**

Using formula (11) we have

$$B''(x|i,4)B''(x|j,4) = \left[\sum_{s=1}^{m+5} L(x|i,s) B(x|s,1)\right] \left[\sum_{t=1}^{m+5} L(x|j,t) B(x|t,1)\right]$$

$$= \sum_{s=1}^{m+5} L(x|i,s) L(x|j,s) [B(x|s,1)]^2$$

$$= \sum_{s=1}^{m+5} L(x|i,s) L(x|j,s) [B(x|s,1)]. \tag{13}$$

This is because $B(x|s,1)$ is an indicator variable of an interval and $B(x|s,1)B(x|t,1) \equiv 0$ if $s \neq t$.

Plugging (13) into (5) yields

$$\begin{aligned}
R_{ij} &= \int B''(x|i,4)B''(x|j,4)dx \\
&= \int \sum_{s=1}^{m+5} L(x|i,s)L(x|j,s)[B(x|s,1)]dx \\
&= \sum_{s=1}^{m+5} \int L(x|i,s)L(x|j,s)[B(x|s,1)]dx \\
&= \sum_{s=1}^{m+5} \int_{t(s)}^{t(s+1)} L(x|i,s)L(x|j,s)dx \\
&= \sum_{s=1}^{m+5} Q_{ijs} \, .
\end{aligned} \qquad (14)$$

Formula (12) can now be used to obtain

$$\begin{aligned}
Q_{ijs} &= \int_{t(s)}^{t(s+1)} L(x|i,s)L(x|j,s)dx \\
&= \int_{t(s)}^{t(s+1)} [a_{is}P(x|s)+b_{is}N(x|s)][a_{js}P(x|s)+b_{js}N(x|s)]dx \\
&= a_{is}a_{js} \int_{t(s)}^{t(s+1)} [P(x|s)]^2 dx \\
&+ a_{is}b_{js} \int_{t(s)}^{t(s+1)} [P(x|s)][N(x|s)]dx \\
&+ b_{is}a_{js} \int_{t(s)}^{t(s+1)} [N(x|s)][P(x|s)]dx \\
&+ b_{is}b_{js} \int_{t(s)}^{t(s+1)} [N(x|s)]^2 dx \, .
\end{aligned} \qquad (15)$$

An application of formula (9) yields

$$\int_{t(s)}^{t(s+1)} [P(x|s)]^2 dx = \int_{t(s)}^{t(s+1)} \left[\frac{x-t(s)}{t(s+1)-t(s)}\right]^2 dx \, .$$

Consider the transformation

$$\begin{aligned}
v &= \frac{x-t(s)}{t(s+1)-t(s)} \\
x &= [t(s+1)-t(s)]v + t(s) \\
dx &= [t(s+1)-t(s)]dv \, .
\end{aligned}$$

Application of this transformation yields

$$\int_{t(s)}^{t(s+1)} [P(x|s)]^2 \, dx = [t(s+1) - t(s)] \int_0^1 [v]^2 \, dv$$

$$= [t(s+1) - t(s)] \left[ \frac{v^3}{3} \right]_0^1$$

$$= \frac{[t(s+1) - t(s)]}{3}. \tag{16}$$

An application of formula (9) yields

$$\int_{t(s)}^{t(s+1)} [P(x|s)][N(x|s)] \, dx = \int_{t(s)}^{t(s+1)} \left[ \frac{x - t(s)}{t(s+1) - t(s)} \right] \left[ \frac{t(s+1) - x}{t(s+1) - t(s)} \right] dx.$$

Another application of the above transformation yields

$$\int_{t(s)}^{t(s+1)} [P(x|s)][N(x|s)] \, dx = [t(s+1) - t(s)] \int_0^1 [v][1 - v] \, dv$$

$$= [t(s+1) - t(s)] \int_0^1 [v - v^2] \, dv$$

$$= [t(s+1) - t(s)] \left[ \frac{v^2}{2} - \frac{v^3}{3} \right]_0^1$$

$$= [t(s+1) - t(s)] \left[ \frac{1}{2} - \frac{1}{3} \right]_0^1$$

$$= \frac{[t(s+1) - t(s)]}{6}. \tag{17}$$

Symmetry yields

$$\int_{t(s)}^{t(s+1)} [N(x|s)]^2 \, dx = \frac{[t(s+1) - t(s)]}{3}. \tag{18}$$

Plugging formulas (16), (17) and (18) into formula (15) yields

$$Q_{ijs} = [t(s+1) - t(s)] \left\{ \frac{a_{is} a_{js}}{3} + \frac{a_{is} b_{js}}{6} + \frac{b_{is} a_{js}}{6} + \frac{b_{is} b_{js}}{3} \right\}. \tag{19}$$

## Summary of Formulas

The summary of the formulas are obtained from formulas (5), (7), (12), (14) and (19). They are

$$\int [CS''(x)]^2 dx = \beta' R \beta,$$

$$\beta' = (\beta_1, \ldots, \beta_{m+2})$$

$R =$ Matrix with $ij$ element $R_{ij}$.

$$R_{ij} = \sum_{s=1}^{m+5} Q_{ijs}$$

$$Q_{ijs} = [t(s+1) - t(s)] \left\{ \frac{a_{is} a_{js}}{3} + \frac{a_{is} b_{js}}{6} + \frac{b_{is} a_{js}}{6} + \frac{b_{is} b_{js}}{3} \right\}$$

$$a_{ik} = \begin{cases} 0 & k = 1, \ldots, i-1 \\ c_i & k = i \\ d_i + e_i & k = i+1 \\ f_i & k = i+2 \\ 0 & k = i+3 \\ 0 & k = i+4, \ldots, m+5 \end{cases}$$

$$b_{ik} = \begin{cases} 0 & k = 1, \ldots, i-1 \\ 0 & k = i \\ c_i & k = i+1 \\ d_i + e_i & k = i+2 \\ f_i & k = i+3 \\ 0 & k = i+4, \ldots, m+5 \end{cases}$$

$$c_i = + \frac{6}{[t(i+3) - t(i)][t(i+2) - t(i)]}$$

$$d_i = - \frac{6}{[t(i+3) - t(i)][t(i+3) - t(i+1)]}$$

$$e_i = - \frac{6}{[t(i+4) - t(i+1)][t(i+3) - t(i+1)]}$$

$$f_i = + \frac{6}{[t(i+4) - t(i+1)][t(i+4) - t(i+2)]}$$

$$c_i = 0 \text{ for } i = 1,2$$
$$d_i = 0 \text{ for } i = 1, m+2$$
$$e_i = 0 \text{ for } i = 1, m+2$$
$$f_i = 0 \text{ for } i = m+1, m+2.$$

## 4. Expansion of the R Matrix

In the theory above, the **R** matrix was derived for the liquid part of a single liquid characteristic. In order to use the roughness penalty concept, the R matrix has to be expanded to the dimension of the entire model. The model level roughness penalty term involves and expanded **R** matrix, and is of the form

$$\beta' * R * \beta,$$

where the dimension of the score coefficient vector, $\beta$, is the number of score coefficients in the model.

Consider the case of maximizing divergence where there is no intercept term. In this case, the expanded **R** matrix is a block diagonal matrix, where each block is associated with one characteristic. For a characteristic with $q$ score coefficients, the block is a $q \times q$ matrix. For a characteristic with no liquid component of order four, the block is a zero matrix.

Now, for example, lets consider a characteristic where the first two attributes are order 1, the next 5 attributes are order 4, and the last attribute is order 1. In this case, the number of score coefficients is 2 + (5 + 3) + 1 = 11. The block matrix, associated with this characteristic, is of the form

$$\begin{bmatrix} 2\times 2 & 2\times 8 & 2\times 1 \\ 0 & 0 & 0 \\ 8\times 2 & 8\times 8 & 8\times 1 \\ 0 & \mathbf{R} & 0 \\ 1\times 2 & 1\times 8 & 1\times 1 \\ 0 & 0 & 0 \end{bmatrix},$$

where the $8 \times 8$ **R** matrix, in the middle, is the matrix discussed in Section 3.

This way, the fitted score coefficients associated with the discrete attributes are not affected by the roughness penalty.

## 5. Max divergence quadratic program

According to References [3] and [6], the max divergence quadratic program without the roughness penalty term is

Find $\boldsymbol{\beta}$ to

Minimize $\boldsymbol{\beta}' * \mathbf{C} * \boldsymbol{\beta} + \dfrac{2\lambda}{n} \boldsymbol{\beta}' * \boldsymbol{\beta}$

Subject to:
  $\mathbf{d}' * \boldsymbol{\beta} = \delta$
  Score Engineering.

In this zero in-weighting case, a simple scale transformation puts you on a weight of evidence scale without disturbing the score engineering. With this formulation, the **H** matrix that goes into the MATLAB Quadratic Programming function is

$$\mathbf{H} = 2\left(\mathbf{C} + \dfrac{2\lambda}{n}\mathbf{I}\right).$$

The max divergence quadratic program with the roughness penalty term is

Find $\boldsymbol{\beta}$ to

Minimize $\boldsymbol{\beta}' * \mathbf{C} * \boldsymbol{\beta} + \frac{2\lambda}{n} \boldsymbol{\beta}' * \boldsymbol{\beta} + \lambda_2 \boldsymbol{\beta}' * \mathbf{R} * \boldsymbol{\beta}$

Subject to:
$\mathbf{d}' * \boldsymbol{\beta} = \delta$
Score Engineering.

With this formulation, the **H** matrix that goes into the MATLAB Quadratic Programming function is

$$\mathbf{H} = 2\left(\mathbf{C} + \frac{2\lambda}{n}\mathbf{I} + \lambda_2 \mathbf{R}\right).$$

Here, the **R** matrix is the expanded **R** matrix discussed in Section 4.

## 6. MATLAB Code

### New MATLAB functions

### CharRP

To implement the roughness penalty in INFORMedge, I developed two new MATLAB functions. The first one is called by the code

R=CharRP(knots,liquid,torder,nsc).

This function computes the block matrix, **R**, associated with the roughness penalty term for a single characteristic with a liquid component.

The inputs are:

knots: The row vector of knots for the characteristic

liquid: The 1 X 2 row vector of knots, which define the liquid range

torder: The derived order of the characteristic

nsc: The number of score coefficients in the characteristic

This **R** matrix has zeros on the perimeter of the matrix to account for the discrete score coefficients that might be associated with the characteristic.

### ModelRP

The second one is called by the code

    R=ModelRP(CHse,model,modstich)

This function computes the matrix, **R**, associated with the roughness penalty term, for the whole model. The penalty term is of the form

$$\beta' * R * \beta,$$

where **R** is a block diagonal matrix. Each block is associated with a characteristic. A block associated with a discrete characteristic is all zeros.

**Modifications to INLPfit**

To run these analyses, I made a small modification to the MATLAB function, INLPfit, and called it INLPfitRP. The calling code is

    score=INLPfitRP(model,algorithm,outcome,Xsc,modstich,statistic, …
            constraint,score,RPM) .

This function has an additional input, RPM, which stands for the Roughness Penalty Matrix. This is the matrix computed by the new function, ModelRP.

So far, I have implemented the roughness penalty capability only in the zero in-weighing max divergence case. For this case, the H matrix for the fitting quadratic program is

    H=2*(C+(2*lambda/n)*eye(p)+RPM).

It would be very easy to implement the roughness penalty capability for the other INLPfit objective functions. I will wait for it to catch on with my fans.

**Fraud analysis**

The data, used for the fraud analysis in Section 7 of this paper, is the MATLAB file, fraud_case.mat.

The inputs for the fraud analysis are in the file, fraud_inputs.m .

The analysis script is in the file, fraud_analysisRP.m.

## 7. Case Study

**Data**

To illustrate the theory, I apply it to models based on the fraud data featured in References [1], [3], [4], [5]. The Table below gives the MATLAB characteristic index and the characteristic name. Unfortunately, I could not uncover from my papers the actual English descriptions of the characteristics. Kimi Minnick. may have them.

**Characteristic indices and names**

| Characteristic Index | Characteristic Name |
| --- | --- |
| 1 | Char170 |
| 2 | Char191 |
| 3 | Char192 |
| 4 | Char200 |
| 5 | Char211 |
| 6 | Char225 |
| 7 | Char314 |
| 8 | Char320 |
| 9 | Char340 |
| 10 | Char380 |
| 11 | Char391 |
| 12 | Char534 |
| 13 | Char607 |
| 14 | Char658 |
| 15 | Char665 |
| 16 | Char706 |
| 17 | Char767 |
| 18 | Char843 |
| 19 | Char901 |
| 20 | Char922 |
| 21 | Char936 |
| 22 | Char950 |
| 23 | Char961 |
| 24 | Char963 |
| 25 | Char965 |

**One characteristic models with Char965 (characteristic index = 25)**

I first explore the effect of the roughness penalty on a set of one characteristic models based on Char965 ( index 25). The knots for this characteristic are
[-2950 -950 -750 -550 -400 -300 -200 -100 80 1425 Inf]. These knots were used in the original INFORM*PLUS* analysis of this data. The liquid range is
[-2950 1425]. For the order 4 case (cubic splines), the number of liquid score coefficients is 9 + 3 = 12. Since there is one discrete attribute ( [1425 Inf) ), the total number of score coefficients is 13.

In the Section 3 theory, the $\mathbf{R}$ matrix is a $12 \times 12$ matrix, because it only involves the liquid score coefficients. But the dimension of $\boldsymbol{\beta}$ for fitting Char965 is $13 \times 1$, because it includes the score coefficient associated with the last discrete attribute.

The model **R** matrix is a $13 \times 13$ matrix of the form

$$\mathbf{R} \leftarrow \begin{bmatrix} \underset{\mathbf{R}}{12 \times 12} & \underset{\mathbf{0}}{12 \times 1} \\ \underset{\mathbf{0}}{1 \times 12} & \underset{\mathbf{0}}{1 \times 1} \end{bmatrix}.$$

**Results with an ascending pattern**

Here are the Characteristic 965 score functions for various values of the smoothness parameter, $\lambda_2$. These functions maximize penalized divergence subject to an ascending pattern over the liquid range. The validation divergence for each model is given in the title of the graph

$\lambda_2 = 0$, **ascending pattern, validation divergence = .296**

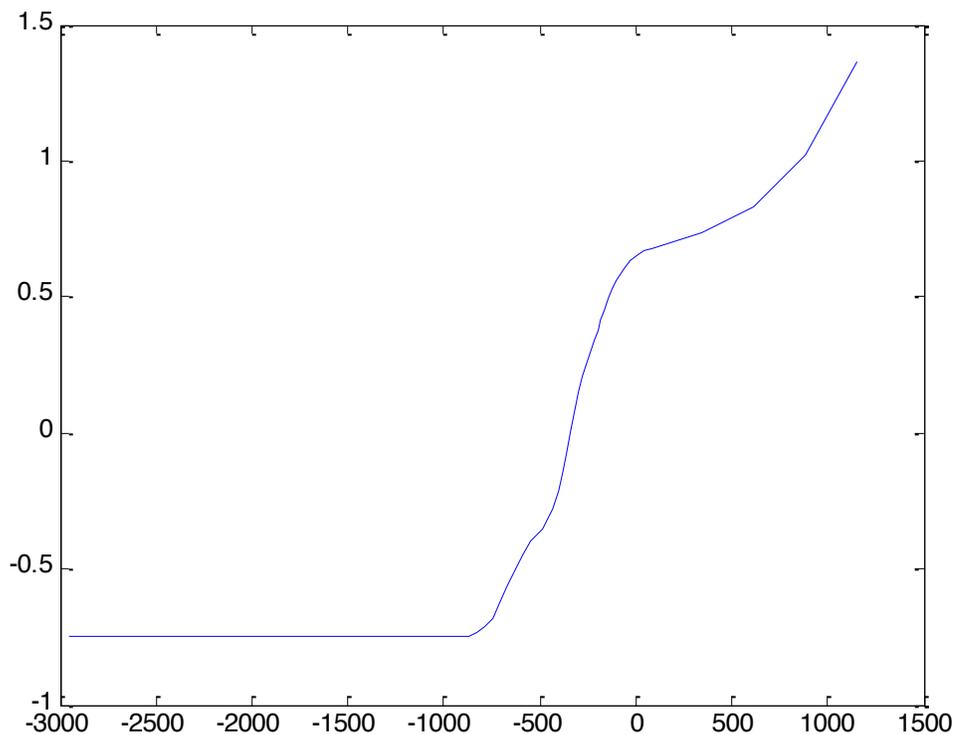

$\lambda_2 = 10^5$, **ascending pattern, validation divergence = .297**

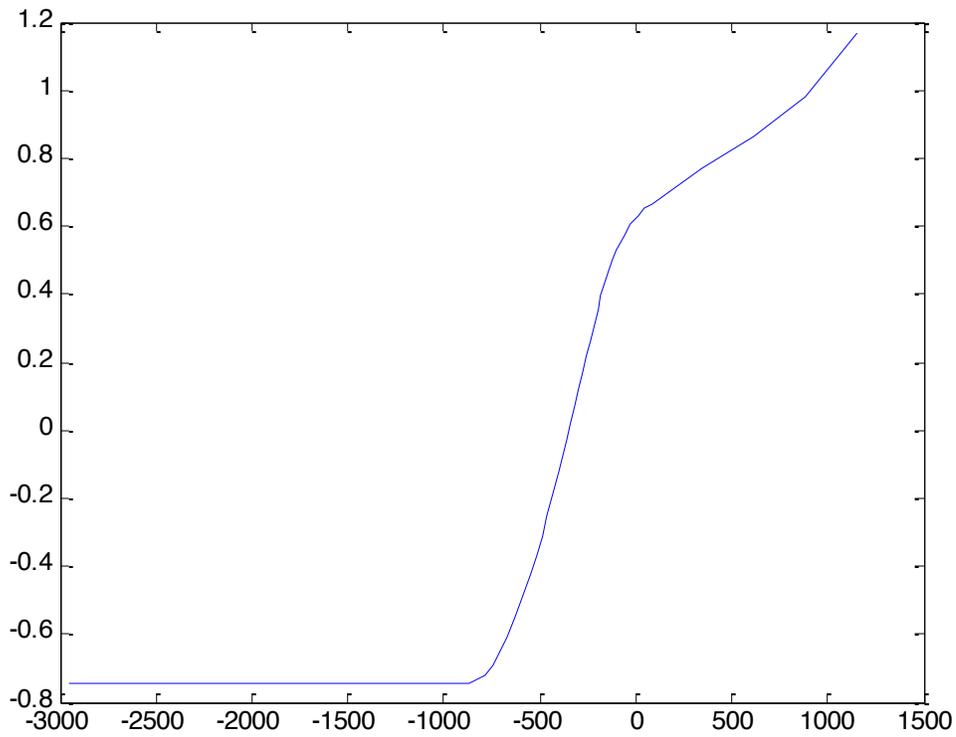

$\lambda_2 = 10^7$, **ascending pattern, validation divergence = .291**

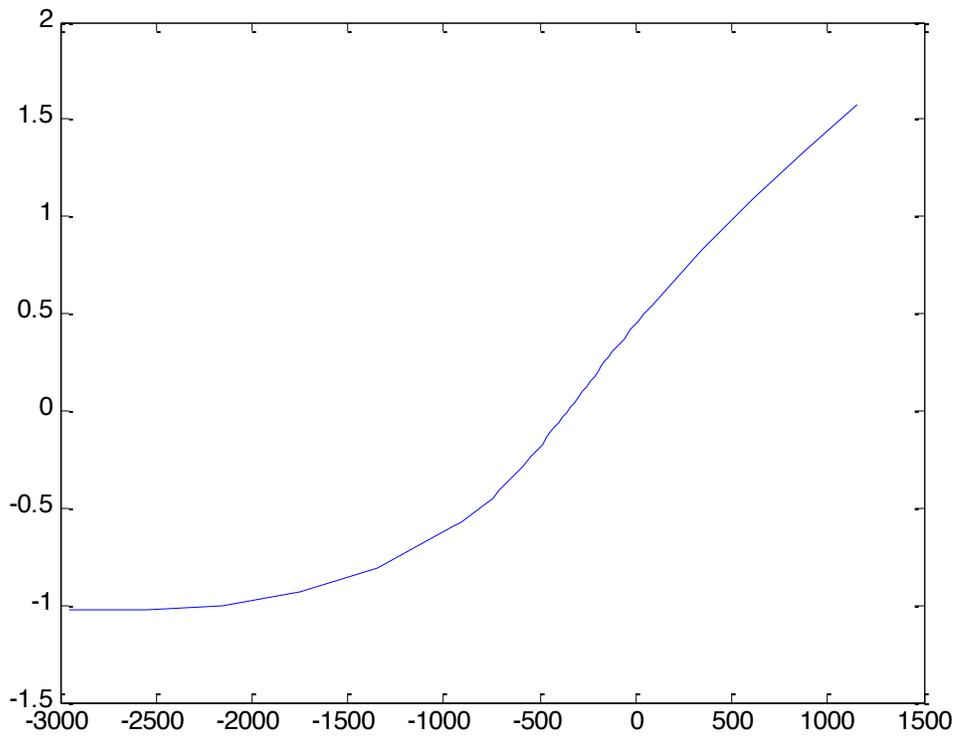

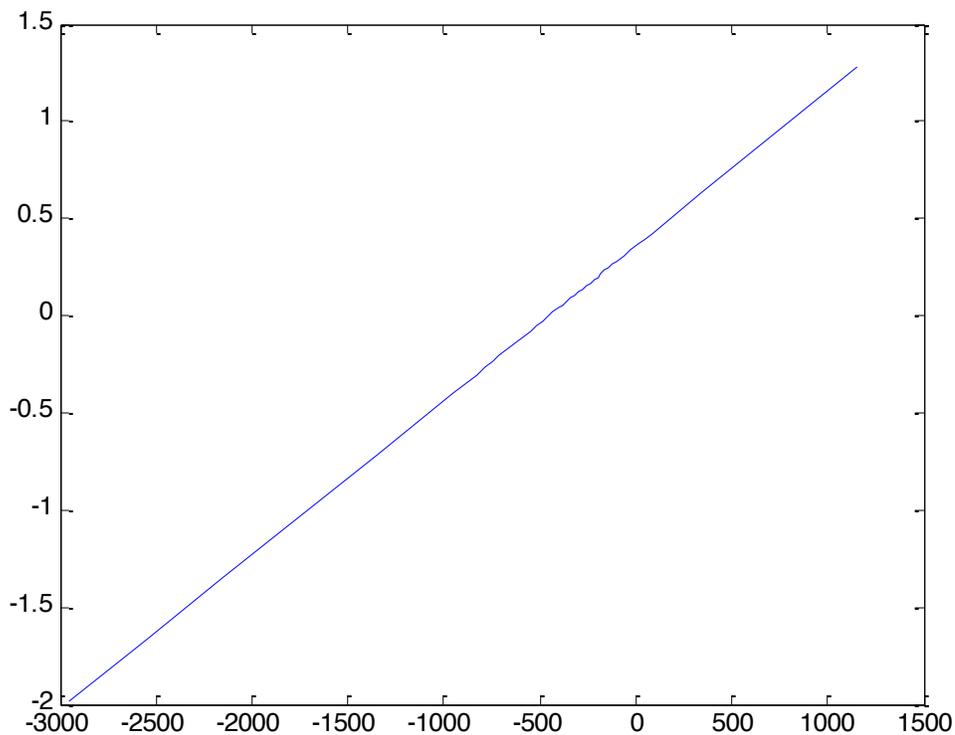

$\lambda_2 = 10^{10}$, **ascending pattern, validation divergence = .255**

For $\lambda_2 = 0$, the characteristic score function has a few wiggles that look spurious. For $\lambda_2 = 10^5$, the characteristic score function looks about right. It is smooth, but retains most of the original shape. For $\lambda_2 = 10^7$, the characteristic score function is very smooth, but deviates from the original shape. For $\lambda_2 = 10^{10}$, the characteristic score function is completely smooth; i.e., linear. For a linear characteristic score, the roughness penalty term is zero, which is as small as it can get. This is because the second derivative of a linear function is zero.

In this example, the validation divergence is maximized at $\lambda_2 = 10^5$, which is also the most palatable model.

**Results with no pattern**

In the graphs above, for $\lambda_2 = 0$, and $\lambda_2 = 10^5$, the characteristic score function has a long plateau from -3,000 to -1,000. This is a sure sign that the data is inconsistent with the ascending pattern. So now I repeat the analysis with the ascending pattern constraint removed.

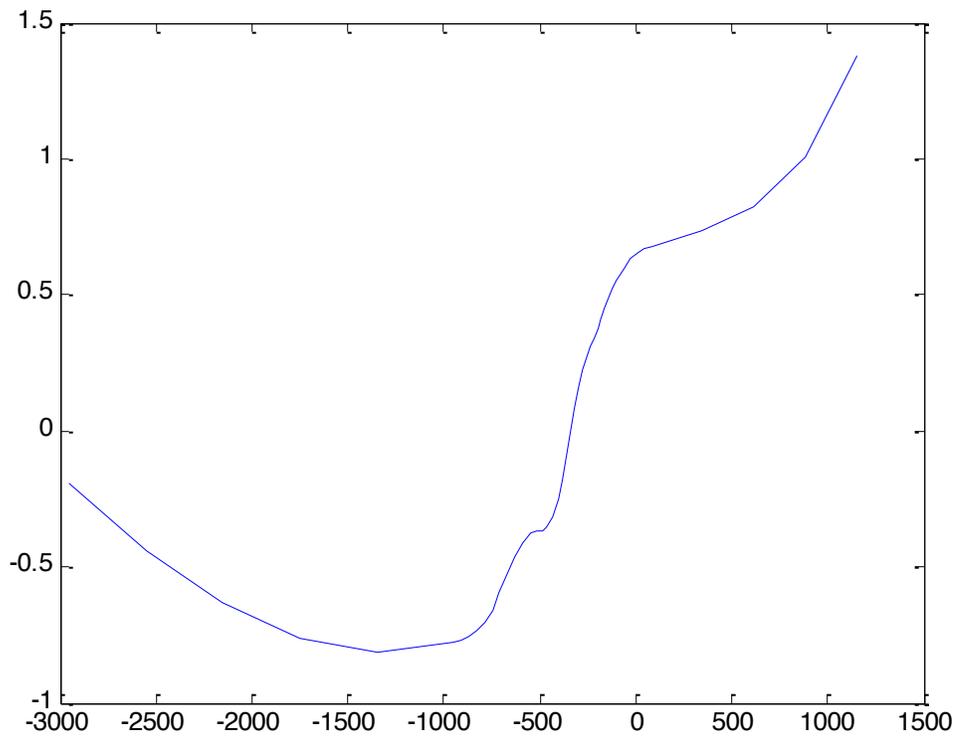

$\lambda_2 = 0$, **no pattern, validation divergence = .292**

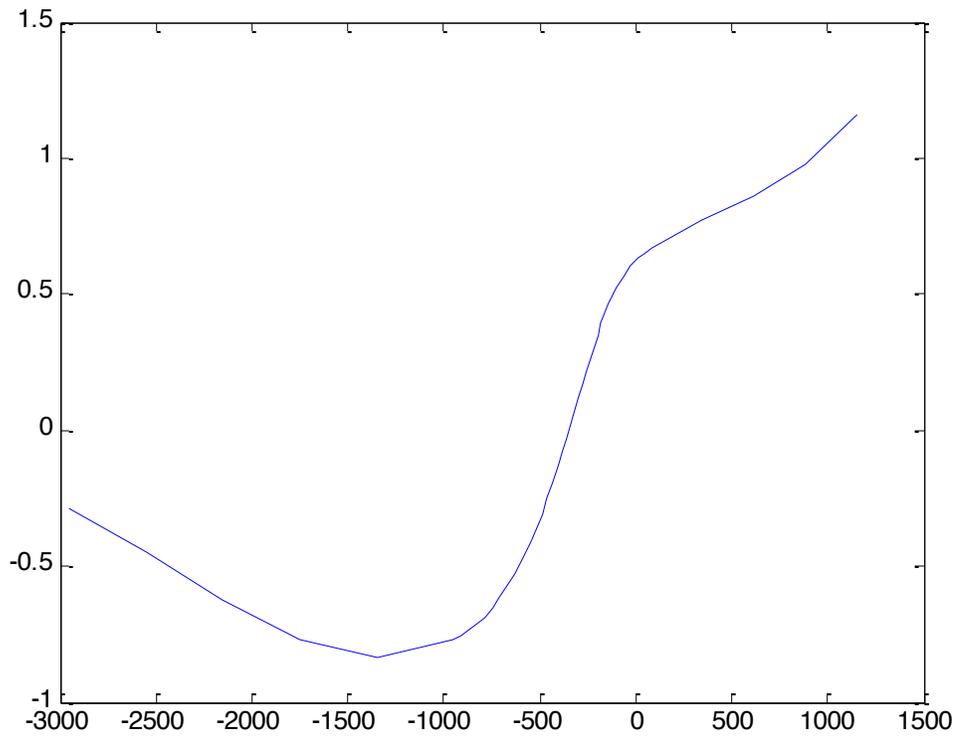

$\lambda_2 = 10^5$, **no pattern, validation divergence = .294**

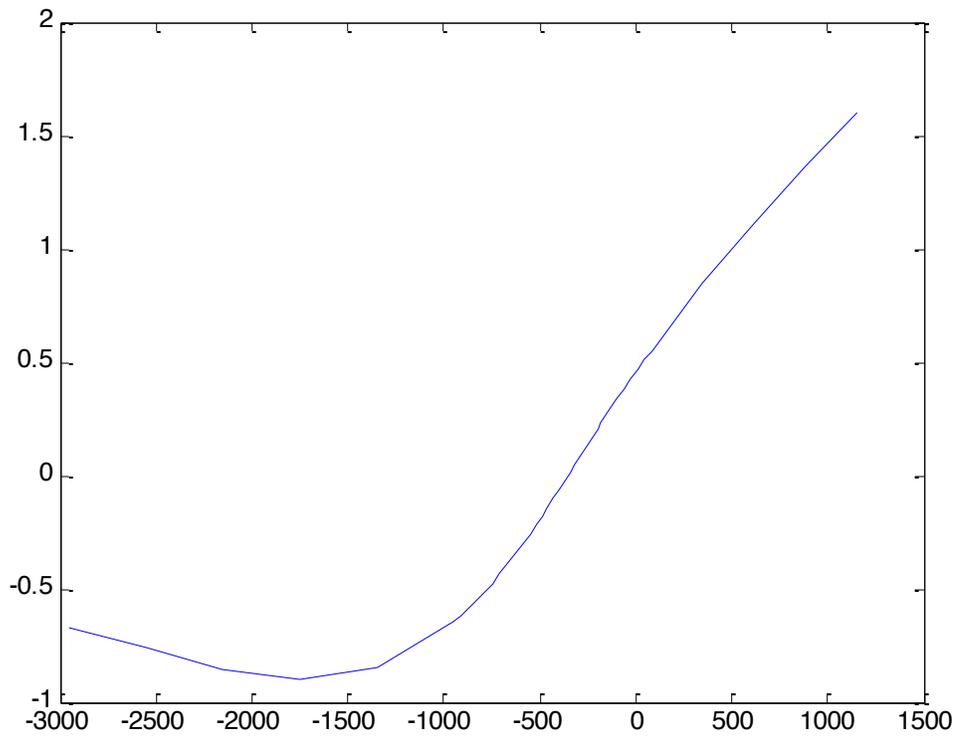

$\lambda_2 = 10^7$, **no pattern, validation divergence = .292**

$\lambda_2 = 10^{7.5}$, **no pattern, validation divergence = .285**

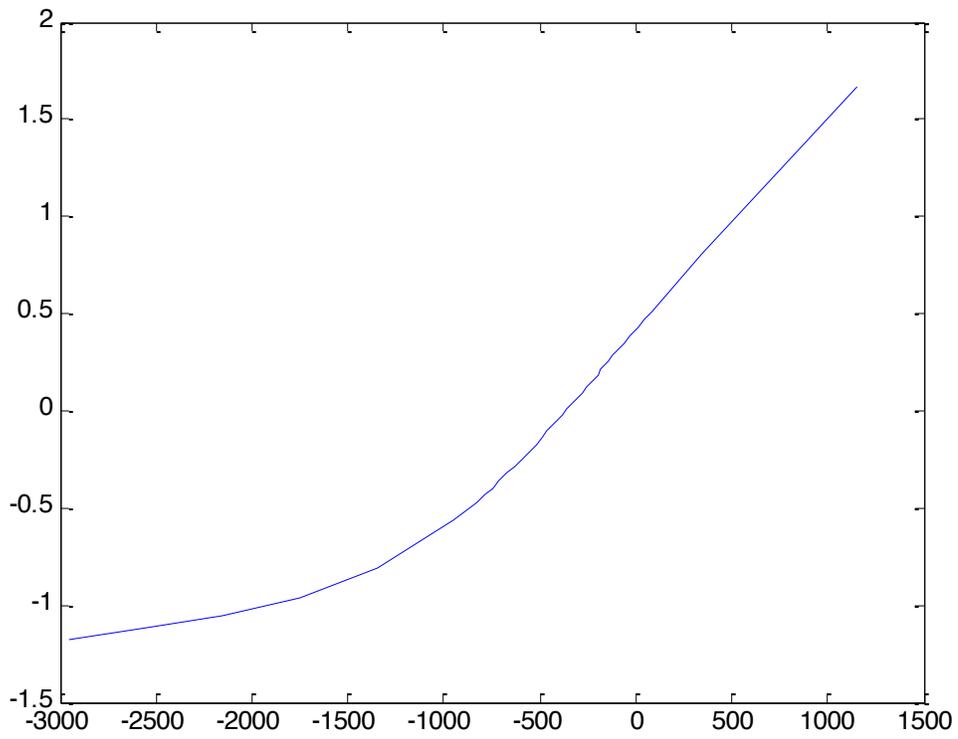

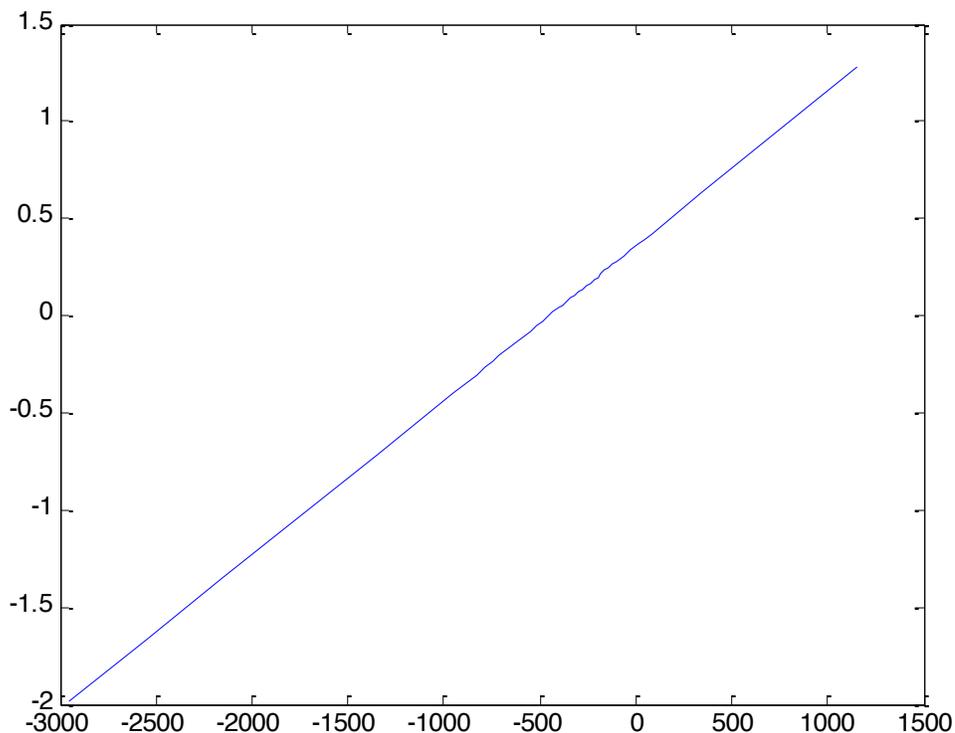

$\lambda_2 = 10^{10}$, **no pattern, validation divergence = .255**

For $\lambda_2 = 0$, the characteristic score function has several wiggles that look spurious. For $\lambda_2 = 10^5$, the characteristic score function is much smoother, but has the same non-monotonicity as the first curve. For $\lambda_2 = 10^7$, the characteristic score function is very smooth, but still not monotonic. For $\lambda_2 = 10^{7.5}$, the characteristic score function is very smooth, and is now monotonic. So this is another way to achieve monotonicity without using a direct pattern constraint. For $\lambda_2 = 10^{10}$, the characteristic score function is completely smooth; i.e., linear.

Note that from a validation divergence point of view, the direct ascending pattern constraint is a better method to achieve an ascending pattern than the smoothness parameter.

**Increasing the number of knots**

The liquid knots for Char965 in the above analyses are

[-2950 -950 -750 -550 -400 -300 -200 -100 80 1425].

It is of interest to examine the consequences of tripling the number of liquid knots to

[-2950 -2250 -1650 -950 -880 -820 -750 -680 -620 -550 -500 -450 -400 -360 -330 -300 -260 -230 -200 -160 -130 -100 -40 20 80 500 1000 1425] .

The results for 4 different inputs are

$$\lambda_2 = 0, \text{ \textbf{no pattern, validation divergence = .299}}$$

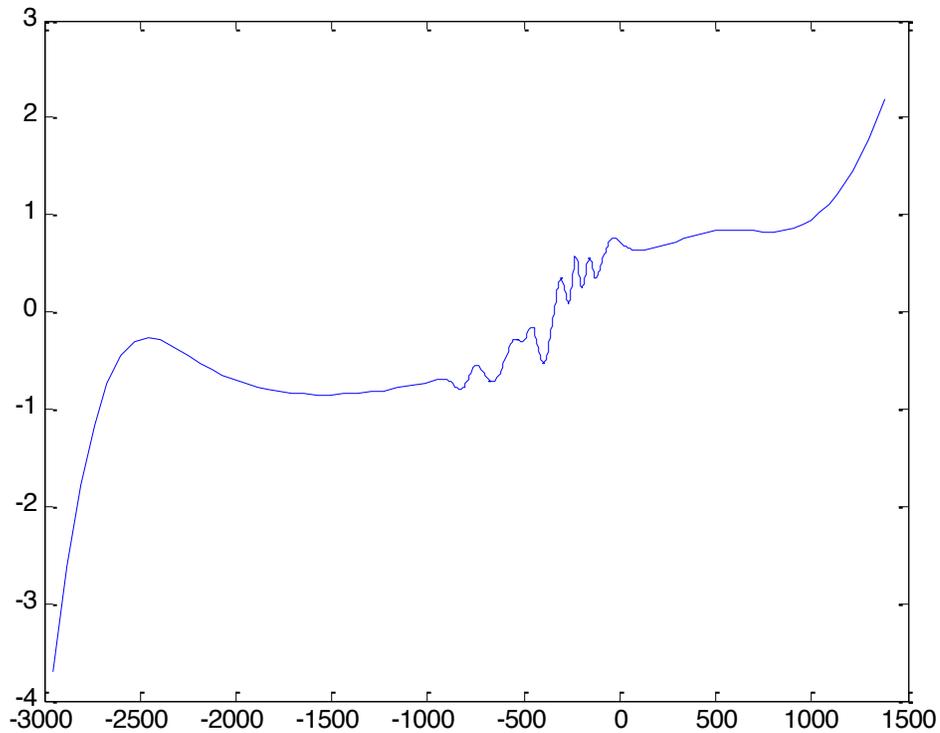

$\lambda_2 = 10^5$, **no pattern, validation divergence = .294**

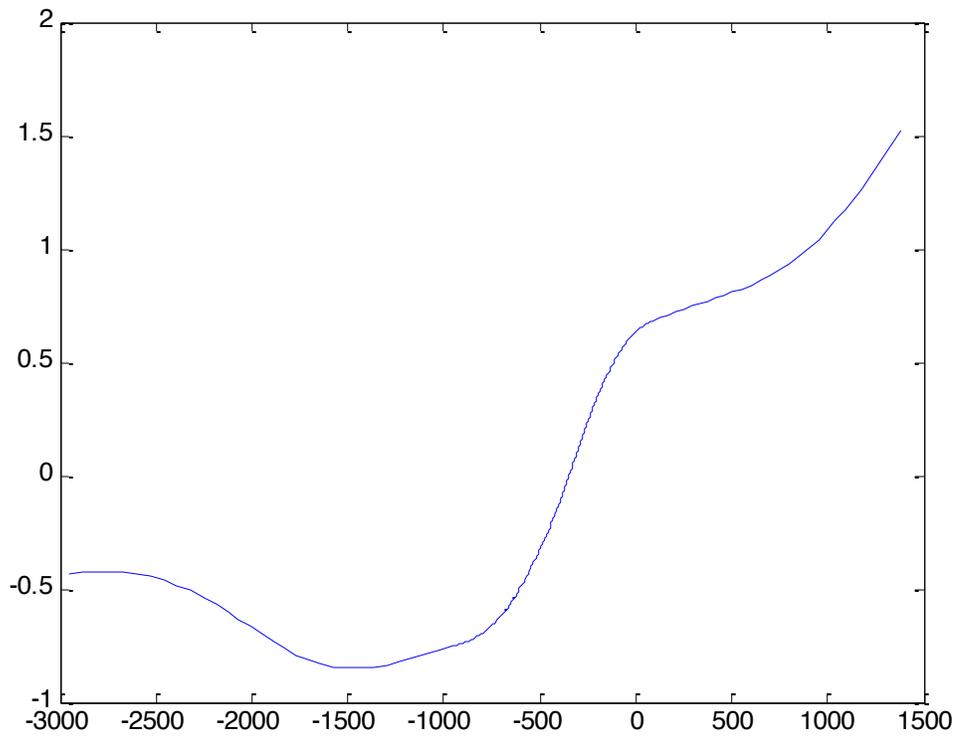

$\lambda_2 = 0$, **ascending pattern, validation divergence = .300**

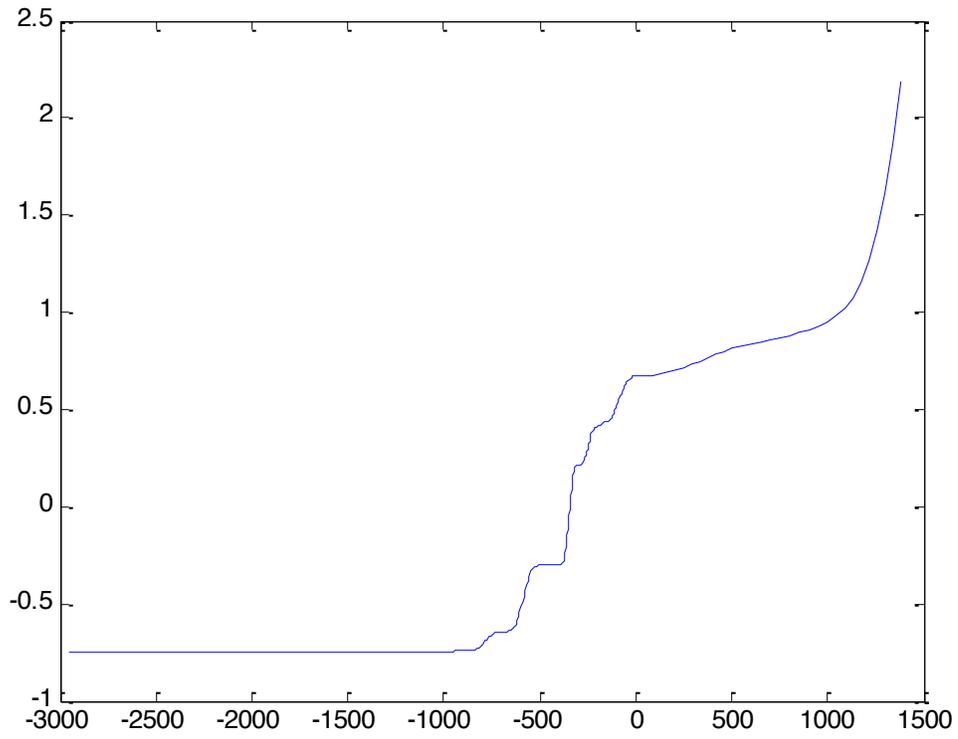

$\lambda_2 = 10^5$, **ascending pattern, validation divergence = .298**

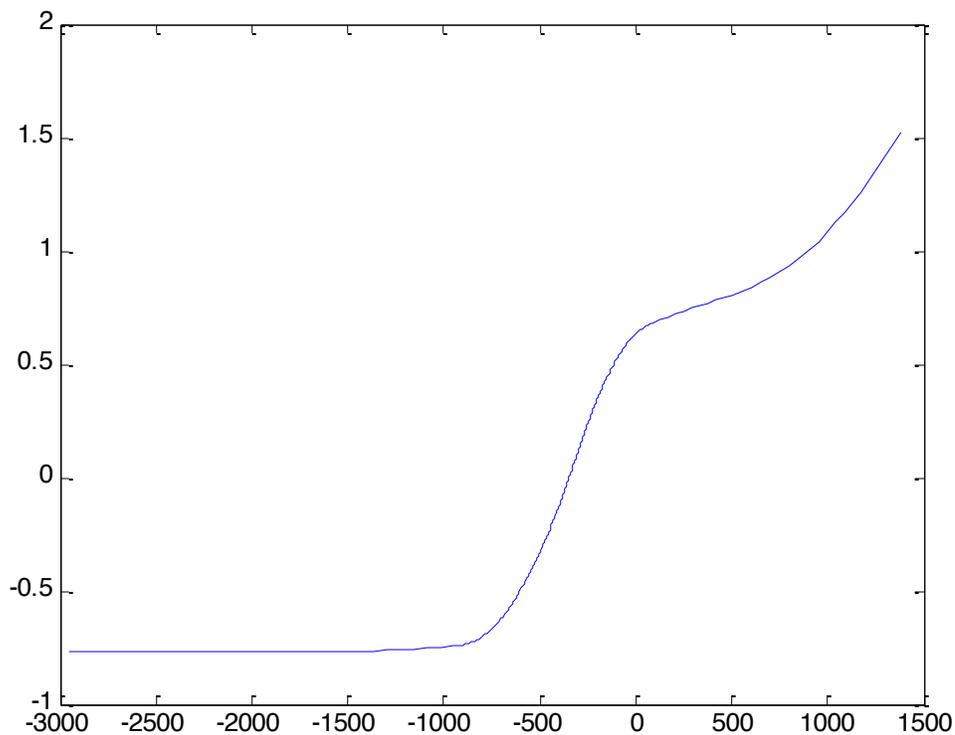

For the case of triple the knots, I think that the most palatable model is the last one, which has an ascending pattern and a significant smoothness parameter. However, this model does not quite maximize the validation divergence.

One explanation for this could be the random error in the validation sample. If this analysis were repeated with other splits of the total data between development and validation, different results might obtain. Another explanation could be that the roughness in the max validation divergence model is caused by correlations with other characteristics that should be in the model, but are not.

**Summary of validation divergences**

Above, I developed many one characteristic models for Char965. Which one is best? One subjective criteria is palatability. For example, the last graph is fairly palatable, but the first graph in the triple knots series is very un-palatable. Another criterion is validation divergence. The table below shows these validation divergences in one place.

**Validation divergence for Char965 models**

| Knots | Smoothness Parameter | Ascending Pattern | Validation Divergence |
|---|---|---|---|
| Original | 0 | Yes | .296 |
| Original | $10^5$ | Yes | .297 |
| Original | $10^7$ | Yes | .291 |
| Original | $10^{10}$ | Yes | .255 |
| Original | 0 | No | .292 |
| Original | $10^5$ | No | .294 |
| Original | $10^7$ | No | .292 |
| Original | $10^{7.5}$ | No | .285 |
| Original | $10^{10}$ | No | .255 |
| Triple the original | 0 | No | .299 |
| Triple the original | $10^5$ | No | .294 |
| Triple the original | 0 | Yes | .300 |
| Triple the original | $10^5$ | Yes | .298 |

Here are a few comments about the results in this table.

1. The ascending pattern constraint improves validation.

2. For the original knots, with the ascending pattern, the best value of the smoothing parameter is $10^5$, which yields a smooth curve with the same overall shape as the curve with the zero smoothing parameter.

3. For the case with triple the original knots, the ascending curve, with maximal validation divergence, has a zero smoothing parameter, and looks a bit rough. This is a little counter-intuitive, because the curve with a smoothing parameter of $10^5$ looks more palatable. However, this particular validation sample is a random sample, and this result might not hold for other independent validation samples.

4. For the case with triple the original knots, no ascending pattern, and zero smoothness parameter, the curve is very wiggly. A reasonable person would surely declare that this curve over-fits. But this declaration is not supported by the validation divergence. Some of these wiggles might be surrogates for a whole bunch of variables, that want to be in the model, but can't, because it is a one characteristic model.

5. My personal favorite is the model with triple the knots, the ascending pattern, and a smoothness parameter of $10^5$ - even though it does not quite maximize validation divergence. Hey, what can I say, I like palatability.

**An issue with the scale of the characteristic**

Another set of one characteristic models is for Char170, which has index 1. The knots for this characteristic are
[-9999999 -9999998 0 5 25 35 300 1000 Inf]. The liquid range is
[0 1000]. For the order 4 case (cubic splines), the number of liquid score coefficients is
5 + 3 = 8. Since there are three discrete attributes, the total number of score coefficients is 11.

I first plot the liquid part of the Char170 score function for smoothness parameter, $\lambda_2 = 0$. This function maximizes penalized divergence subject to a descending pattern over the liquid range. The *x*-axis is actually (Characteristic + 1), because I want to stay away from zero.

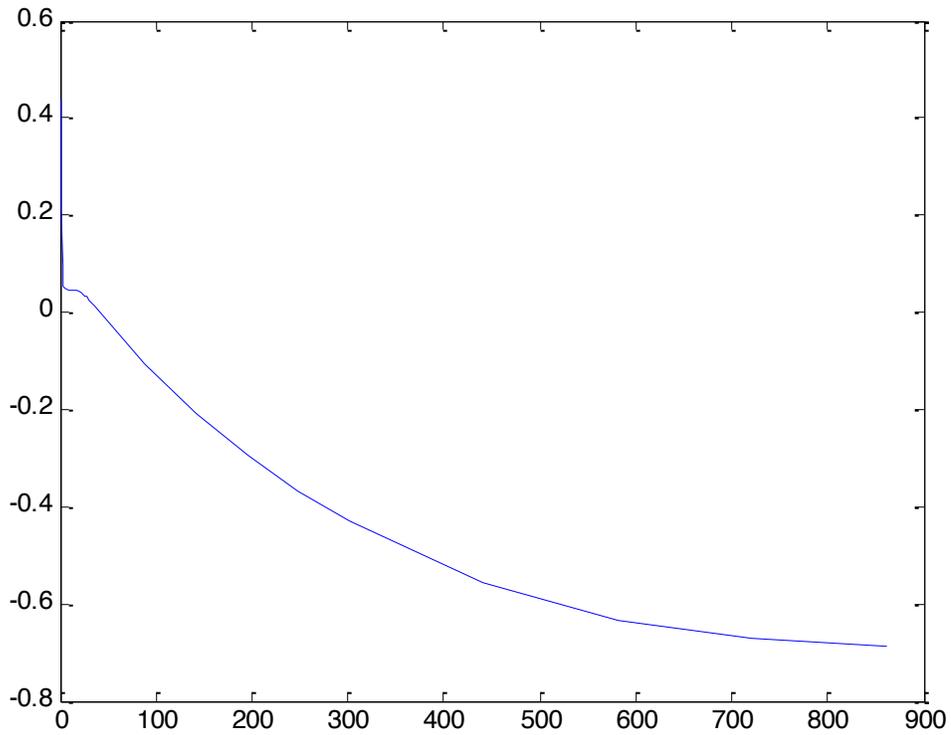

$\lambda_2 = 0$, **Natural scale for $(x + 1)$**

In this curve, the characteristic score drops suddenly from 0.43 to 0.047 as $(x + 1)$ goes from 1 to 5. The early part of the curve looks very rough.

But if we plot the curve on a log scale, we get a very different picture

$$\lambda_2 = 0, \text{ Log scale for } (x + 1)$$

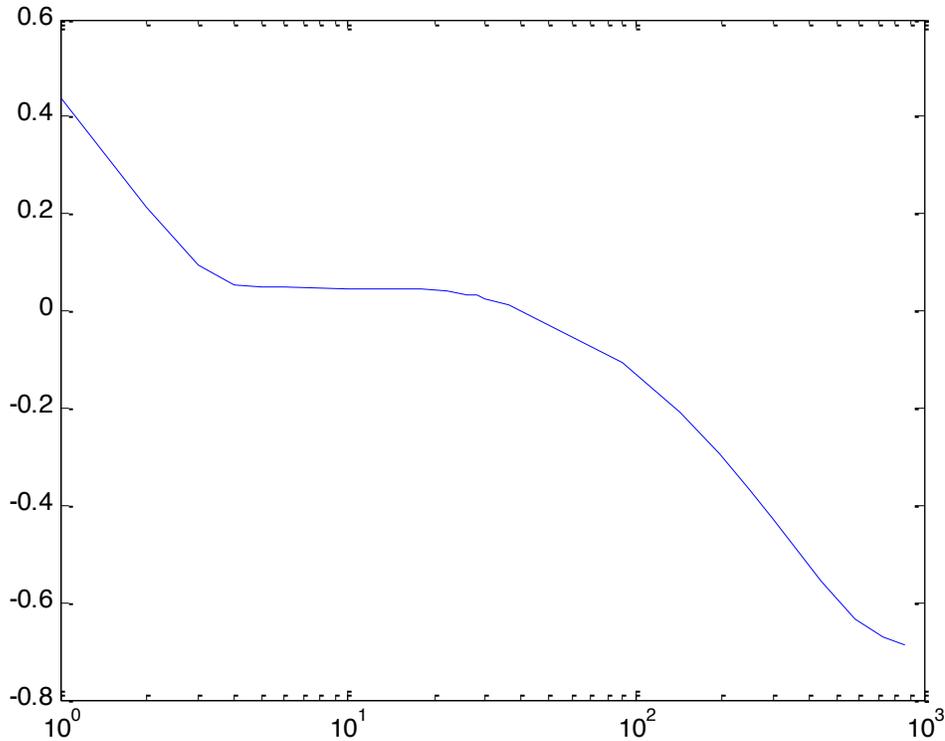

Now the curve looks fairly smooth. What is an analyst to do?

For $\lambda_2 = 10^{2.5}$, the characteristic score functions, for the natural and log scales, look as follows:

$\lambda_2 = 10^{2.5}$, **Natural scale for (*x* + 1)**

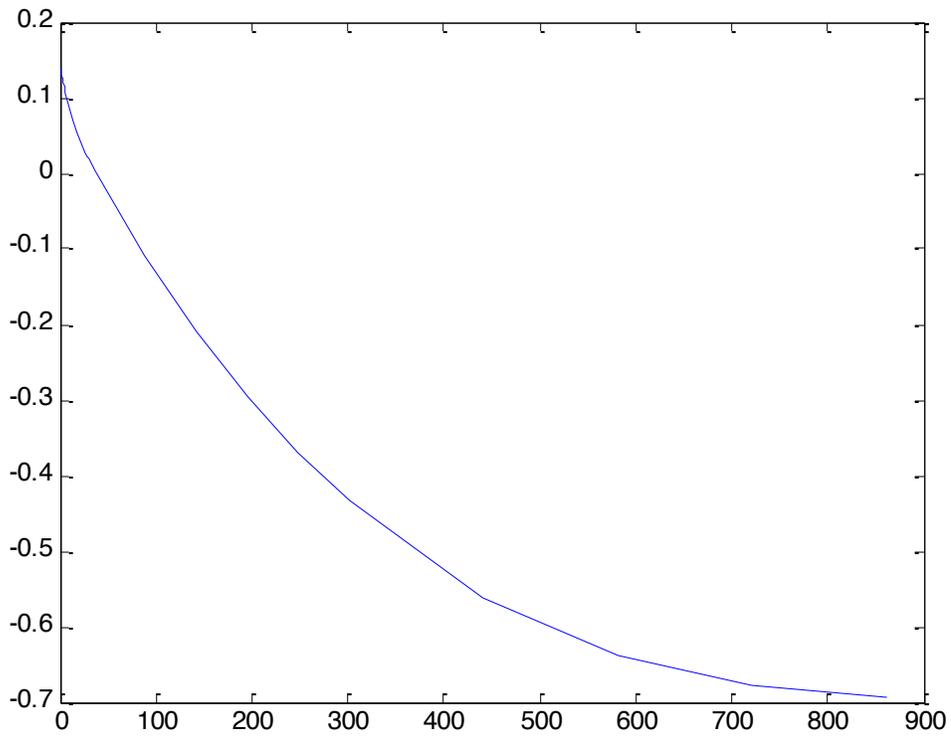

$\lambda_2 = 10^{2.5}$, **Log scale for $x + 1$**

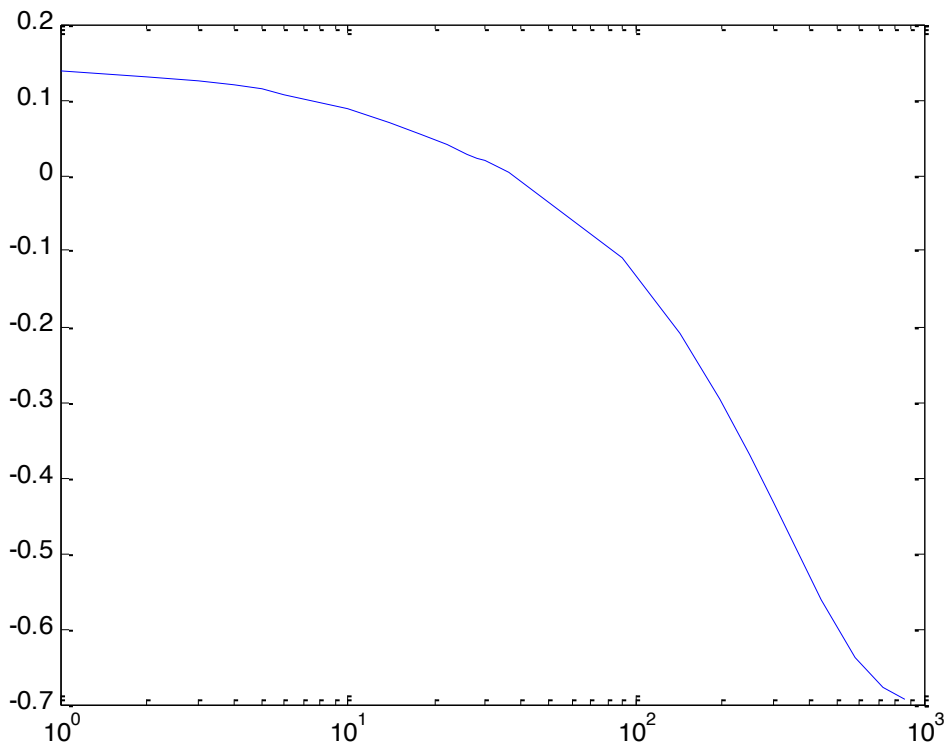

These two curves now look fairly smooth and might satisfy the desire for palatability.

If all you care about is model performance, then you might want to choose a value of the smoothness parameter that maximizes your favorite performance metric on a validation sample.

**Optimizing characteristic level smoothness parameters**

In this analysis, I first developed a score using 12 liquid characteristics – all with zero smoothness parameters. The characteristics used had characteristic indices 1-6, 8, 14, 22-25. The validation divergence was 1.198. I then ordered the characteristics by Step I marginal contribution. Characteristic 4 (Char200) had the maximum Step I marginal contribution. I then found the smoothness parameter value for Characteristic 4, which maximized the validation divergence. It was $\lambda_2 = 10$. I held this constant for the rest of the analysis.

Characteristic 1 (Char170) had the second largest Step I marginal contribution. I then found the smoothness parameter value for Characteristic 1, which maximized the validation divergence. It was $\lambda_2 = 0$, and was held at this value for the rest of the analysis.

In a similar manner, I proceeded through Characteristics 25, 3, 5, 8, 23, 2, 22. The marginal contributions and the "optimal" values of $\lambda_2$ are shown in the following Table. The word optimal is put in quotes, because this was a greedy optimization, which would not have produced the true optimum.

**"Optimal" smoothness parameters**

| Characteristic Index | Characteristic Name | Step I Marginal Contribution | "Optimal" Smoothness Parameter ($\lambda_2$) |
|---|---|---|---|
| 4 | Char200 | .247 | 10 |
| 1 | Char170 | .087 | 0 |
| 25 | Char965 | .080 | 100 |
| 3 | Char192 | .068 | 1 |
| 5 | Char211 | .058 | $10^{3.5}$ |
| 8 | Char320 | .033 | $10^7$ |
| 23 | Char961 | .032 | $10^{7.5}$ |
| 2 | Char191 | .028 | 0 |
| 22 | Char950 | .021 | $10^5$ |
| 6 | Char225 | .017 | - |
| 24 | Char963 | .016 | - |
| 14 | Char658 | .012 | - |

The final validation divergence was 1.206. This is a modest improvement over the original validation divergence of 1.198.

All of the characteristic score functions, for which "optimal" smoothness parameters were computed, are plotted below. Note that a log scale is used for Characteristics 1 through 4.

**Characteristic 1 with** $\lambda_2 = 0$

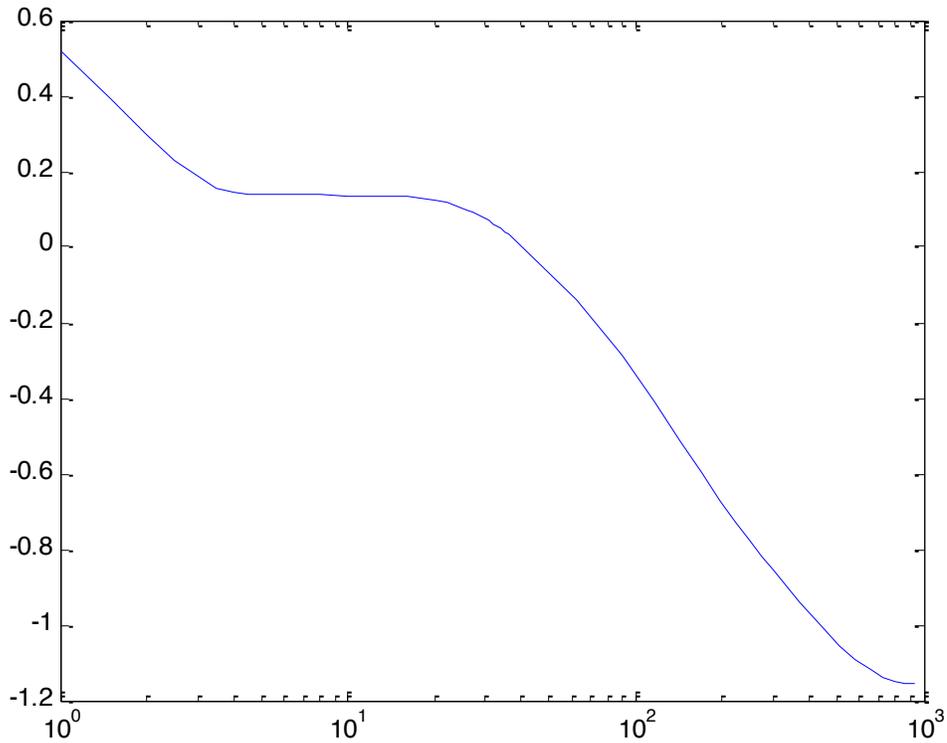

**Characteristic 2 with** $\lambda_2 = 0$

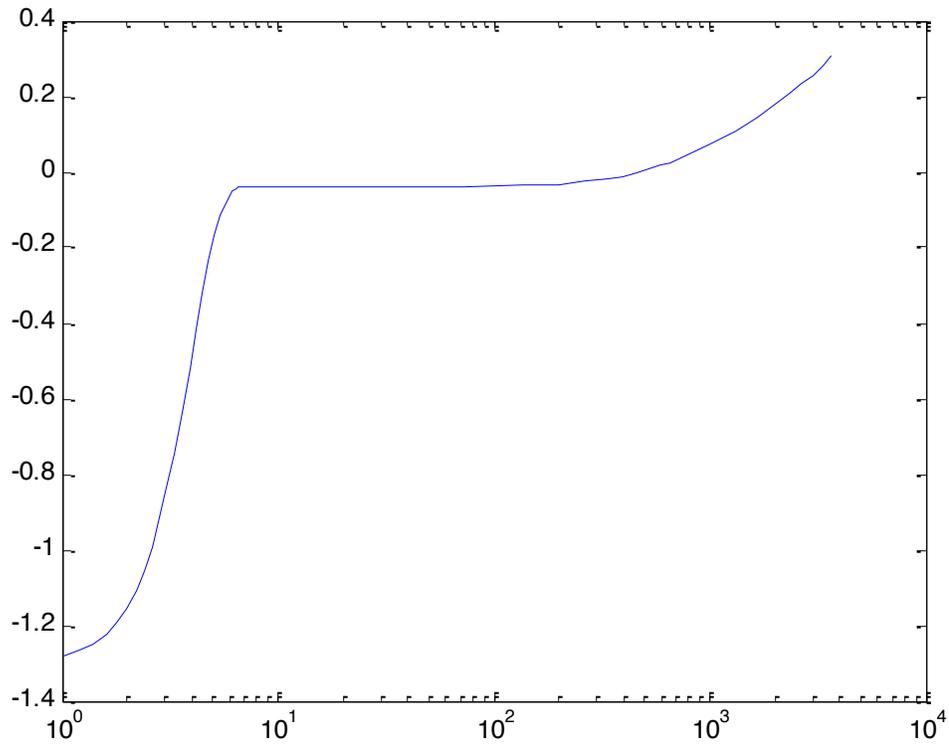

**Characteristic 3 with** $\lambda_2 = 1$

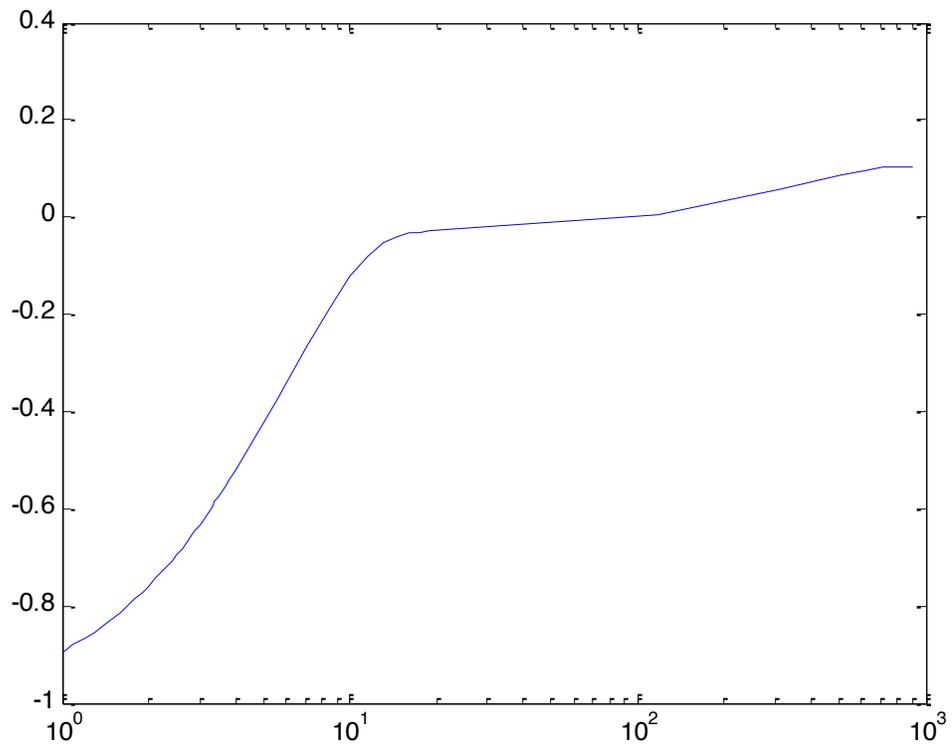

## Characteristic 4 with $\lambda_2 = 10$

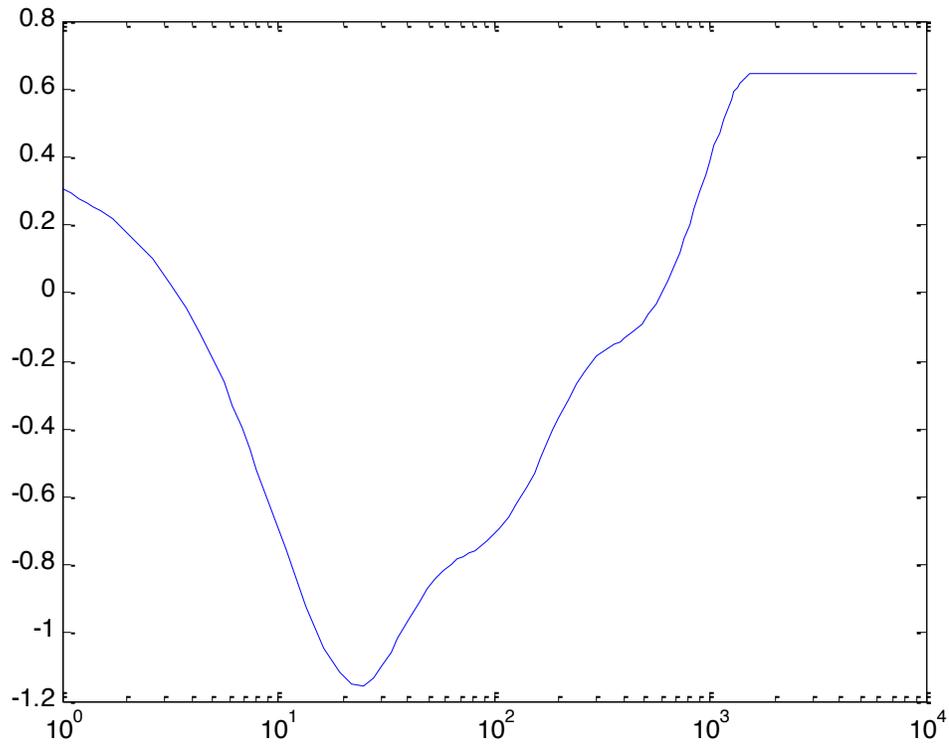

## Characteristic 5 with $\lambda_2 = 10^{3.5}$

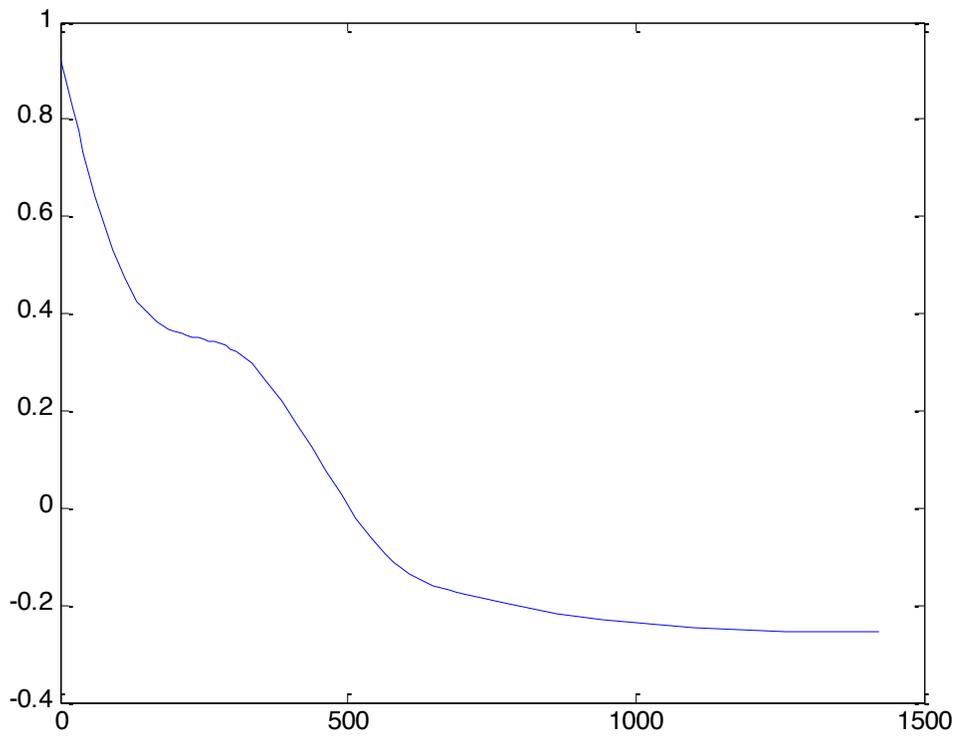

**Characteristic 8 with** $\lambda_2 = 10^7$

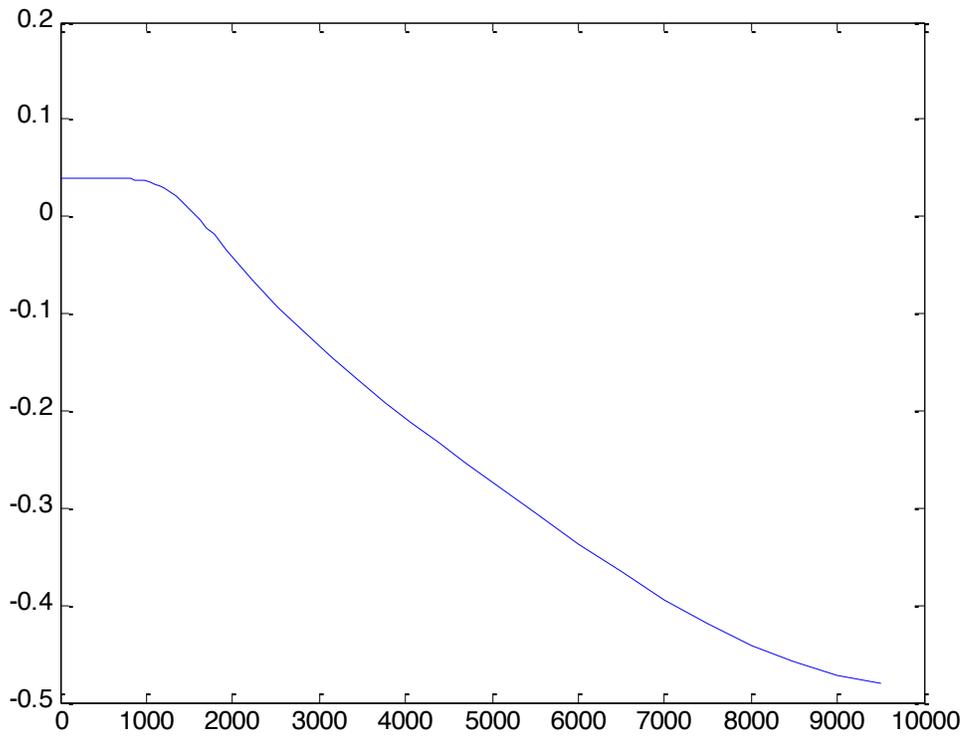

**Characteristic 22 with $\lambda_2 = 10^5$**

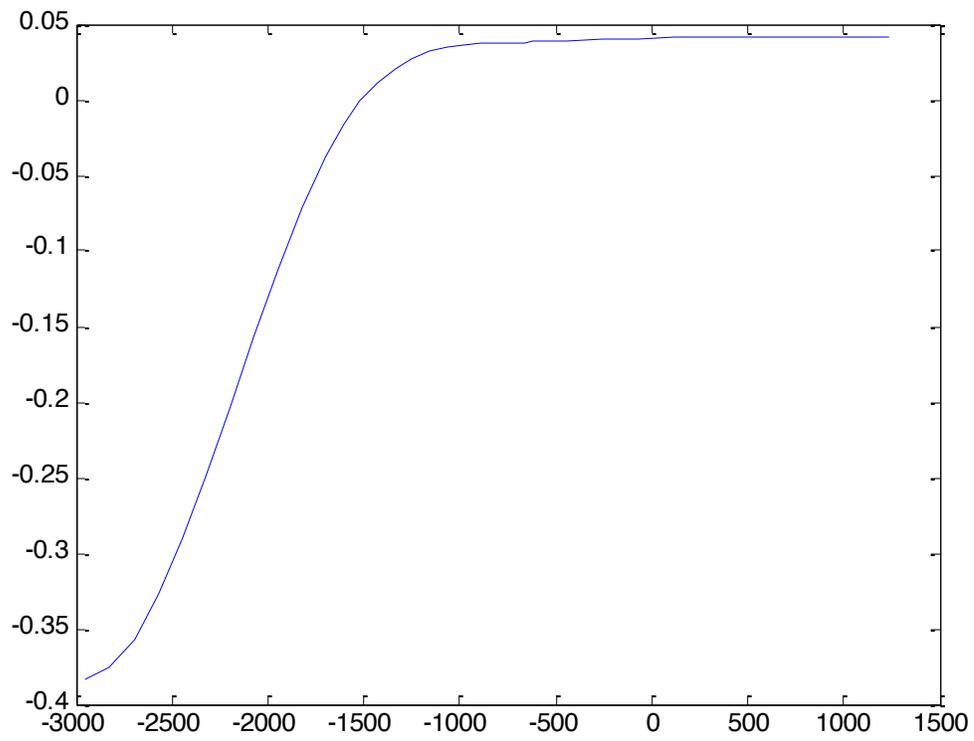

**Characteristic 23 with $\lambda_2 = 10^{7.5}$**

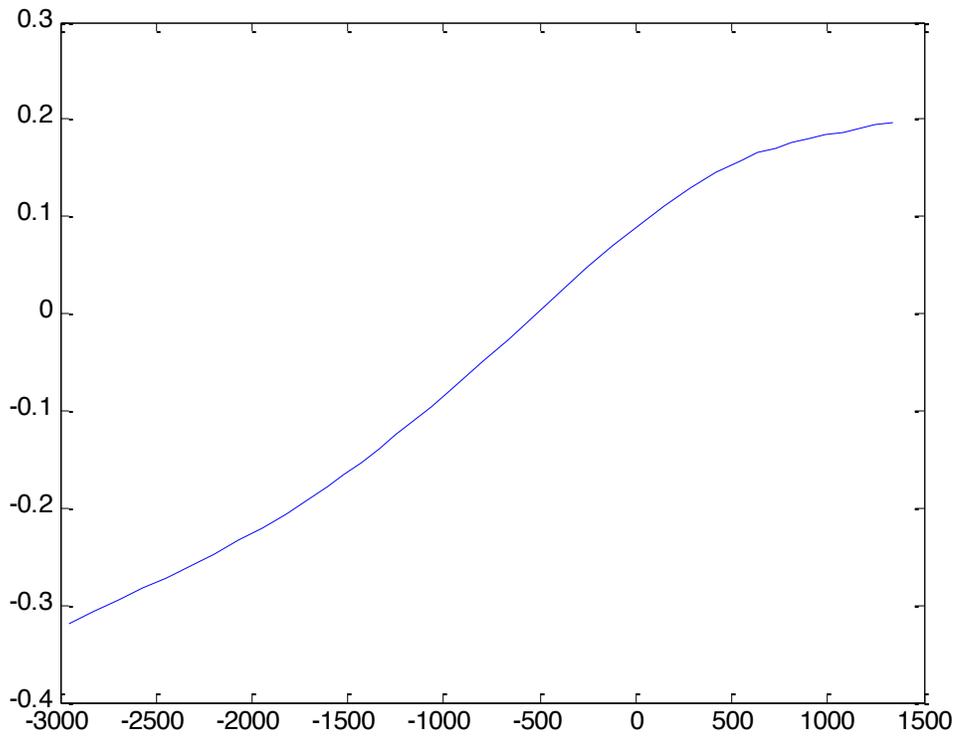

**Characteristic 25 with** $\lambda_2 = 100$

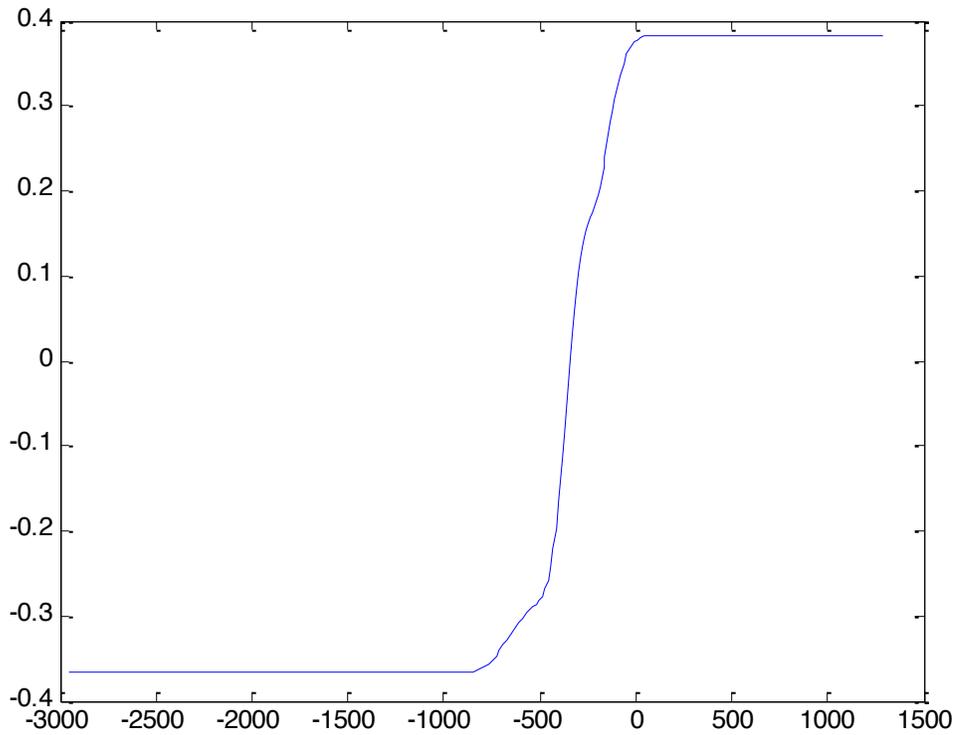

Most of these curves look nice and smooth. The exceptions are Characteristics 2, 4, and 25, which have smoothness parameters 0, 10, and 100 respectively. So the model that maximizes validation divergence, is not the one with all very smooth characteristic scores.

# 8. Other Applications of the Roughness Penalty

**Smoothing a traditional scorecard**

The characteristic scores for a traditional scorecard are step functions. Sometimes, these step functions look rough; i.e., if you draw a curve through the centers of the steps, then this curve will look rough. This visualization suggests how to apply the theory, herein, to smooth out the step function.

Associated with the step functions are the knots defined by the attributes. If the attributes are not numerical, then transform them to some reasonable numerical scale. For this discussion, ignore the weird knots like -9999999 and -9999998.

With these knots on reasonable scales, fit smooth cubic splines to the data using the characteristic level smoothness parameters. The "smooth" step functions would be some kind of step function approximations to these smooth cubic splines.

Here is one method. Let $CS(x)$ be a smooth cubic spline. For a given attribute of a characteristic, the score weight would be the sample weighted average

$$\frac{\sum_{x_i \in \text{Attribute}} sw_i \bullet CS(x_i)}{\sum_{x_i \in \text{Attribute}} sw_i}.$$

**Coarse classing**

In Reference [7], I developed a new method of coarse classing. This method was based on estimating the true characteristic score with a step function. The estimating step function was the one that minimized the average point-wise mean squared error.

This method required that you estimate the true characteristic score with a smooth curve. For testing the theory in Reference [7], Nina Shikaloff used a SAS method of smoothing. The smoothing method in this paper could be used rather than the SAS method, because it is more flexible and powerful.

**Are coarse classes really needed?**

The technical answer to this question is no – at least for the liquid parts of characteristic scores. As demonstrated in this paper, if you use fine classes with a positive smoothing parameter, you will get a smooth characteristic score.

But there are practical considerations that might call for coarse classing. If you fit only with fine classes, then the number of score coefficients in the model might be quite large. This could slow down the fitting algorithm noticeably.

Also, if you fit with fine classes, then you might have to experiment with many smoothing parameters (one for each characteristic) in order to get a reasonable model. Experimenting in such a high dimension could be time consuming. However, you could automate this experimenting with an Object*Boost* (Reference [8]) type algorithm.